\documentclass[conference]{IEEEtran}
\usepackage{amsfonts}

\usepackage{ifpdf}
 \ifpdf
 \else
 \fi

\ifCLASSINFOpdf
   \usepackage[pdftex]{graphicx}
   \graphicspath{{../pdf/}{../jpeg/}}
   \DeclareGraphicsExtensions{.pdf,.jpeg,.png}
\else
   \usepackage[dvips]{graphicx}
   \graphicspath{{../eps/}}
   \DeclareGraphicsExtensions{.eps}
\fi
\usepackage[cmex10]{amsmath}
\usepackage{algorithm}
\usepackage{algorithmic}
\usepackage{array}
\usepackage{mdwmath}
\usepackage{mdwtab}
\usepackage{eqparbox}
\usepackage[tight,footnotesize]{subfigure}
\usepackage[caption=1]{caption}
\usepackage[caption=false,font=footnotesize]{subfig}
\usepackage{stfloats}
\fnbelowfloat
\usepackage{url}
\usepackage{amssymb}
\usepackage{bm} 
\usepackage{float}
\usepackage{indentfirst}
\usepackage{makeidx}
\usepackage{subfigure}
\usepackage{graphicx}
\usepackage{booktabs}
\usepackage{multirow}

\begin{document}
\title{Soft Consistency Reconstruction: A Robust 1-bit Compressive Sensing Algorithm }
\author{\IEEEauthorblockN{Xiao Cai, \IEEEauthorrefmark{2}Zhaoyang Zhang, Huazi Zhang, Chunguang Li}

\IEEEauthorblockA{~Institute of Information and Communication
Engineering, Zhejiang University, Hangzhou 310027, China\\
Zhejiang Provincial Key Laboratory of Information Network Technology, Zhejiang Province, China\\
E-mail: caixiaozju@zju.edu.cn, \IEEEauthorrefmark{2}ning\_ming@zju.edu.cn, hzhang17@zju.edu.cn, cgli@zju.edu.cn}
}

\date{}

\maketitle

\begin{abstract}
A class of recovering algorithms for 1-bit compressive sensing (CS) named Soft Consistency Reconstructions (SCRs) are proposed. Recognizing that CS recovery is essentially an optimization problem, we endeavor to improve the characteristics of the objective function under noisy environments. With a family of re-designed consistency criteria, SCRs achieve remarkable counter-noise performance gain over the existing counterparts, thus acquiring the desired robustness in many real-world applications. The benefits of soft decisions are exemplified through structural analysis of the objective function, with intuition described for better understanding. As expected, through comparisons with existing methods in simulations, SCRs demonstrate preferable robustness against noise  in low signal-to-noise ratio (SNR) regime, while maintaining comparable performance in high SNR regime.
\end{abstract}

\IEEEpeerreviewmaketitle

\section{Introduction}
Compressive sensing (CS), as an emerging signal processing technique, has drawn considerable research interests in recent years, due to its potential to revolutionize future digital communication, wireless networking and even broader areas. Its applications can be found in many fields, such as image recovery \cite{egiazarian2007compressed}, radar detection \cite{baraniuk2007compressiveradar}, spectrum sensing\cite{zhang2011distributed}\cite{10-sw-icc}, channel estimation \cite{bajwa2010compressed} and random access \cite{fazel2012compressed}. Even though, fundamental works on the algorithm design still bear substantial significance and await further breakthroughs.

CS reconstructs sparse signal by specific non-linear algorithms with its sampling rate significantly lower than the Nyquist rate \cite{candes2006compressive}\cite{donoho2006compressed}. An $N$-dimension signal x is defined as $K$-sparse if it satisfies $||x||_0\doteq |supp(x)|\leqslant K$. The signal is sampled into $M$ measurements by measurement matrix $\mathbf{\Phi }\in \mathfrak{R}^{M\times N}$,
\begin{equation}
y=\mathbf{\Phi }x+n,
\end{equation}
where $n$ is the $N$-dimension noise vector. It has been demonstrated in \cite{candes2008restricted} that $\mathbf{\Phi }$ obeying restricted isometry property (RIP) guarantees accurate recovery of the sparse signal $x$ with high probability.

In practice, we can never acquire infinitely precise $y$. Real-valued measurements are quantized to discrete bits. The extreme 1-bit quantization puts forward the notion of 1-bit compressive sensing \cite{boufounos20081}. The value of each measurement is confined to a binary output.
\begin{equation}
y=sign(\mathbf{\Phi }x+n),
\end{equation}
where $sign(\cdot)$ is the sign function, equaling $1$ for positive and $-1$ for negative.


After the introduction of  1-bit compressive sensing, several recovery algorithms have been developed. The goal is to search for the best sparse estimate from the candidate space. Binary Iterative Hard Thresholding (BIHT) algorithms \cite{jacques2011robust} outperform the other similar algorithms, such as Matching Sign Pursuit (MSP) \cite{boufounos2009greedy} and Restricted Step Shrinkage (RSS) \cite{laska2011trust}. It is shown that the one-sided $l_2$-norm (BIHT-$l_2$) objective penalizes on the overall error, and thus excels in the advent of frequently switching measurements caused by noise. Alternatively, the one-sided $l_1$-norm (BIHT-$l_1$) objective stringently forces measurement consistency between the original and the estimated signal, thus is suitable for less noisy scenarios with fewer sign flips in measurements..

On the one hand, high level of noise at the quantizer will flip the near-zero positive measurements to negative (and vice versa), which is the main and the most common cause of performance deterioration during reconstruction. On the other hand, noisy measurements are very common in all types of CS applications. Hence, the counter-noise issue deserves to be addressed as a supplement to the existing literature. To the best of our knowledge, \cite{yan2012robust} provides a method to deal with the sign flips of noisy measurements. It picks out positions where sign flip may occur and recovers the sparse signal from ``seemingly correct" measurements, outperforming BIHTs significantly. Unfortunately, the best performance is achieved using prior knowledge, i.e., the exact number of distorted measurements. Therefore, the method has some limitations.

In this paper, we propose a class of CS recovery algorithms named Soft Consistency Reconstructions (SCR). The contributions of this paper are twofold.
\begin{enumerate}
\item We re-design the objective function in the CS recovery algorithm using soft-decision-based consistency criterion, and provide structural analysis to justify its superiority over the conventional counterparts.

\item We develop a class of SCR algorithms which, in the absence of prior knowledge on noise distortion, work well in both noisy and noiseless scenarios, and compare their performances with the existing algorithms, i.e., BIHTs.
\end{enumerate}

Three aspects are taken into consideration while comparing SCRs with BIHTs, as summarized in Table \ref{SBcomparison}.

\begin{table}[h]
  \centering
  \caption{Comparison between SCRs and BIHTs}
    \begin{tabular}{|c|c|c|}
    \hline
    \multicolumn{2}{|c|}{Prior knowledge} & SCRs=BIHTs \\
    \hline
    \multicolumn{2}{|c|}{Complexity} & SCRs=BIHTs \\
    \hline
    \multirow{2}[4]{*}{Accuracy} & high SNR & SCRs$\approx$BIHTs \\
\cline{2-3}          & low SNR & SCRs$>$BIHTs \\
    \hline
    \end{tabular}%
    \label{SBcomparison}
\end{table}%

The remainder of this paper is organized as follows. Section I reviews the problem formulation of BIHT algorithms. In Section III, the notion of SCR is introduced and the algorithms are described with their advantages analyzed. Section IV validates the proposed algorithms through extensive simulations. Finally, conclusions are drawn in section V.

\section{A Revisit of the BIHT Algorithms}
Consistency is crucial for CS reconstruction algorithms. In BIHTs, the purpose is to find the sparse candidate vector (or estimate), which, after left multiplied by the measurement matrix $\Phi$, has the most elements being consistent with the measured vector $y$. Let $\phi_i$ be the i-th row of the measurement matrix $\mathbf{\Phi }$, and the above description of consistency requires
\begin{equation}
y_i = sign(\phi_i \hat{x})
\end{equation}
\begin{equation}\label{consistency}
\Leftrightarrow y_i\cdot (\phi_i \hat{x}) \geqslant 0.
\end{equation}

Therefore, the BIHT algorithms in \cite{jacques2011robust} (BIHT-$l_1$ or BIHT-$l_2$) are designed based on the following optimization problem
\begin{align}\label{op1}
\mathop {\min }\limits_{x\in \mathfrak{R}^N}& \sum_{i=1}^{M}D(y_i\phi_i x)+\lambda||x||_1,\\
\text{s.t.}&\quad ||x||_2^2=1,\quad\quad
\nonumber
\end{align}
where $D(\cdot)$ is the one-sided $l_1$ (or $l_2$)-norm:
\begin{equation}
D(t)=\left\{\begin{matrix}
0, & if\quad t\geqslant 0\\
-t~(or~t^2), & if\quad t<0
\end{matrix}\right.
.
\end{equation}

Different choices of the parameter $\lambda$ will result in different sparsity of the estimate $\hat{x}$. When the sparsity $K$ of $x$ is unavailable, we can set $\lambda$ according to how many largest coefficients in $x$ we are interested in. There exist some proper $\lambda$s that can achieve the exact sparsity $K$. In this case, it is most likely to find a very accurate estimate. So the reconstruction problem of (\ref{op1}) evolves to
\begin{align}\label{op2}
\mathop {\min }\limits_{x\in \mathfrak{R}^N}& \sum_{i=1}^{M}D(y_i\phi_i x),\\
\text{s.t.}&\quad ||x||_2^2=1, ||x||_0 \leqslant K.
\nonumber
\end{align}

\section{Soft Consistency Reconstruction Algorithms}
In this section, we introduce the novel idea of Soft Consistent Reconstruction (SCR). We first describe the inspirations behind the new consistency metric, and then provide some performance analysis in comparison with the conventional methods. Finally, we formalize the SCR algorithms and explain the roles of its two key parameters.

\subsection{A Novel Consistency Metric}
\begin{figure}[!h]
\centering
\includegraphics [width=3in]{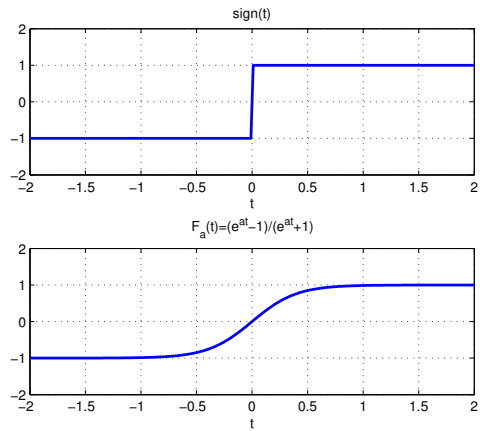}
\caption{$sign(t)$ v.s. $F_a(t)$}
\label{sign}
\end{figure}

First of all, let us examine the following function $F_a(\cdot)$
\begin{equation}
F_a(t)= \frac{e^{at}-1}{e^{at}+1},
\end{equation}
where the parameter $a$ reflects the steepness of $F_a(t)$ near the origin (e.g., a=5 in Fig. \ref{sign}).

Designed to imitate the binary output of the 1-bit quantizer in the encoder, the function $F_a(t)$ is used by the decoder to produce a soft output, also ranging from -1 to 1, from the estimated real-valued measurements $\mathbf{\Phi }\hat{x}$. In soft consistency reconstructions, we measure consistency through calculating the difference between $F_a(\phi_i\hat{x})$ and its corresponding 1-bit measurement, namely $|F_a(\phi_i \hat{x})-y_i|$. As in Fig. \ref{sign}, $|F_a(\phi_i \hat{x})-y_i|$ will approach zero only if the absolute value of $\phi_i\hat{x}$ is larger than $1/a$ and has the same sign as $y_i$, indicating consistency. A $\phi_i\hat{x}$ with the same sign as $y_i$ but having a small absolute value will still result in a noticeable value in $|F_a(\phi_i\hat{x})-y_i|$.

Therefore, to minimize the overall inconsistency of M items, the general optimization problem is expressed using a novel objective
\begin{align}\label{op3}
\mathop {\min }\limits_{x\in \mathfrak{R}^N}& \sum_{i=1}^{M}|y_i-F_a(\phi_i x)|^p,\\
\text{s.t.}&\quad ||x||_2^2=1, ||x||_0 \leqslant K,
\nonumber
\end{align}
where the parameter $p$ denotes power of the inconsistency term. Different $p$s produce different forms of SCR algorithms, such as SCR-1, SCR-2, SCR-3 and so on, collectively called the SCR family.

\emph{Remarks:} An intuitive explanation of adopting the current form of $F_a(t)$ is that, as will be elaborated later, $F_a(t)$ is able to mitigate the flipping effect of noise on near-zero measurements, compared with the conventional consistency metric of $sign(t)$.

\subsection{Structural Analysis}
To provide some insights into the differences between BIHTs and SCRs, we convert (\ref{op3}) into a similar form of (\ref{op2}).
\begin{align}\label{op4}
\mathop {\min }\limits_{x\in \mathfrak{R}^N}& \sum_{i=1}^{M}G_a(y_i\phi_i x)^p,\\
\text{s.t.}&\quad ||x||_2^2=1, ||x||_0 \leqslant K,
\nonumber
\end{align}
where function $G_a(\cdot)$ is a variant of $F_a(\cdot)$
\begin{equation}\label{definitionG}
G_a(t)= 1-F_a(t).
\end{equation}

The equivalence is demonstrated as follows.

\begin{proof}
To demonstrate that (\ref{op3}) and (\ref{op4}) are equivalent, it suffices to prove that
\begin{equation}\label{equi}
|y_i-F_a(\phi_i x)|=G_a(y_i\phi_i x)
\end{equation}

The function $F_a(\cdot)$ is a centrally symmetric function, that is,
\begin{equation}\label{central}
F_a(-t)=\frac{e^{-at}-1}{e^{-at}+1}=\frac{1-e^{at}}{1+e^{at}}=-F_a(t)
\end{equation}

If $y_i=1$, considering $|F_a(\phi x)|\leqslant 1$ and (\ref{definitionG}), there exists
\begin{equation}\label{y=1}
|y_i-F_a(\phi_i x)|=1-F_a(y_i\phi_i x)=G_a(y_i\phi_i x).
\end{equation}

If $y_i=-1$, considering $|F_a(\phi x)|\leqslant 1$, (\ref{definitionG}) and (\ref{central}), there exists
\begin{align}\label{y=-1}
&|y_i-F_a(\phi_i x)|=1+F_a(\phi_i x)\\
\nonumber
=&1-F_a(y_i\phi_i x)=G_a(y_i\phi_i x).
\nonumber
\end{align}

Therefore, combining (\ref{y=1}) and (\ref{y=-1}), we have (\ref{equi}).
\end{proof}

\begin{figure} [htb]
\centering
\includegraphics [width=3in]{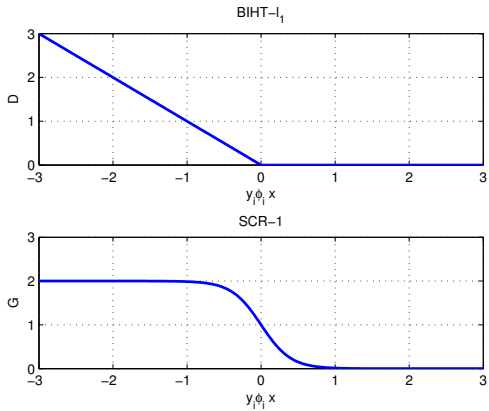}
\caption{$D(y_i\phi_i x)$ in BIHT-$l_1$ vs. $G(y_i\phi_i x)$ in SCR-$1$}
\label{G}
\end{figure}

After the transformation, the term $y_i\phi_i x$ should be paid attention to, which is used to calculate consistency in both (\ref{op2}) (BIHTs) and (\ref{op4}) (SCRs). A negative $y_i\phi_i x$ indicates inconsistency. The larger the absolute value of the negative $y_i\phi_i x$ has, the lower the level of consistency is. Because (\ref{op2}) and (\ref{op4}) are both functions of $y_i\phi_i x$, the essential difference between BIHTs and SCRs lies clearly in the distinct structures of $D(\cdot)$ and $G(\cdot)$, as presented in the Fig. \ref{G}.

In Fig. \ref{G}, we put BIHT-$l_1$ and SCR-1 together. In a structural analysis of the two functions $D(\cdot)$ and $G(\cdot)$, one similarity and one difference should be paid attention to. The similarity is that they are both monotonically decreasing functions and their values approach zero when $y_i\phi_i x\rightarrow\infty$. In this case, gradient descent method will promote a larger $y_i\phi_i x$ until the gradient approximates zero.

The difference lies in the right part of curves. For all $0<y_i\phi_i x<y_j\phi_j x$, we have
\begin{align}
\textsl{BIHT-$l_1$}: &D(y_i\phi_i x)=D(y_j\phi_j x)=0,\label{17}\\
\textsl{SCR-1}: &G(y_i\phi_i x)>G(y_j\phi_j x)>0.\label{18}
\end{align}

\emph{Remarks:} (\ref{17}) means that SCR-1 still has a significant impact on positive $y_i\phi_i x$ near zero. It believes that in noisy environment, a small positive $y_i\phi_i x$ is likely to be flipped from a negative $y_i\phi_i x$ due to positive noise, and thus has low credibility to guarantee the correctness of the  corresponding $y_i$. By contrast, according to (\ref{18}), BIHT-$l_1$ only aims to push $y_i\phi_i x$ above zero and does not think so much.

Simply put, the SCR algorithms try to seek consistency between the estimate $\hat{x}$ and the original signal $x$, while the BIHT algorithms aim at forcing consistency between the estimate $\hat{x}$ and the binary quantization output $y$.

Note that it is easy to  take the derivative of (\ref{op4}), therefore the optimization problem can be solved by gradient descent methods, as is shown below.

\subsection{Soft Consistency Reconstruction Algorithm}
The proposed algorithm searches for the optimal solution to the optimization problem of (\ref{op4}). The steps are straightforward, as shown in Algorithm 1. In the interest of space, the derivation process of Step 3 is given in Appendix.

\begin{algorithm}[!h]
\caption{Soft Consistency Reconstruction}
\label{algorithm}
\begin{algorithmic}
\STATE \textbf{Inputs:} $y\in\{\pm1\}^M$, $\mathbf{\Phi }$.
\STATE \textbf{Initialization:}  $l=0$, $l_{max}$,  $a$, $p$, $K$, $\tau$,
\STATE $\quad\quad\quad\quad\quad\quad \hat{x}^0=\frac{\mathbf{\Phi}^T y}{||\mathbf{\Phi}^T y||}$.
\STATE $\quad$ \textbf{Repeat}
\STATE 1. $\quad l\leftarrow l+1$
\STATE 2. $\quad g(i)\leftarrow G_a(y_i\phi_i\hat{x}^{l-1})$, $i=1,2,...,M$.
\STATE 3. $\quad t\leftarrow -\frac{1}{2}ap\mathbf{\Phi }^T [y\odot g^p\odot(2-g)]$.
\STATE 4. $\quad b\leftarrow \hat{x}^{l-1}-\tau\cdot t$.
\STATE 5. $\quad u\leftarrow b|_K$.
\STATE 6. $\quad \hat{x}^l=\frac{u}{||u||_2}$.
\STATE $\quad$\textbf{until} $l=l_{max}$.
\STATE $\quad$\textbf{return} $\hat{x}^l$
\end{algorithmic}
\end{algorithm}

Notations and procedure descriptions: In the algorithm,  $g$ is an M-dimension column vector and Step 2 calculates its elements one by one. The operator $\odot$ in Step 3 represents the element-wise multiplication. The operator $(\cdot)|_K$ keeps K largest elements of a vector and sets the other elements to zero.

The algorithm is originally designed for the case where sparsity $K$ is available. However, the algorithm can still work when K is unavailable, given that we gradually force a sparse solution. In other words, we can replace Step 5 with the following
\begin{equation}
u\leftarrow sign(b)\cdot\max(|b|-\tau\lambda\cdot\mathbf{1},\mathbf{0}).
\end{equation}

Then the modified algorithm tries to find the solution to an optimization problem similar to (\ref{op1})
\begin{align}\label{op5}
\mathop {\min }\limits_{x\in \mathfrak{R}^N}& \sum_{i=1}^{M}G_a(y_i\phi_i x)^p+\lambda||x||_1,\\
\text{s.t.}&\quad ||x||_2^2=1.
\nonumber
\end{align}

The steps of SCR algorithm resemble those of BIHT.  The key difference lies in the calculation of gradient in Step 3. Therefore, the SCR algorithm brings in no additional complexity than BIHT.

\begin{figure} [!h]
\centering
\includegraphics [width=3in]{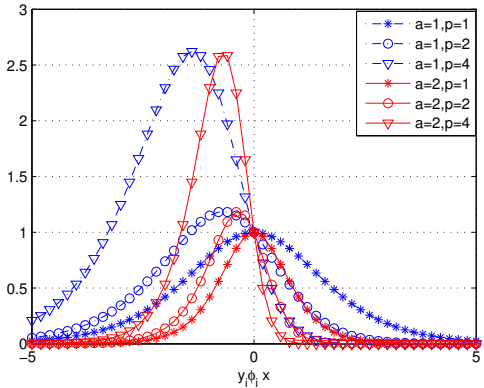}
\caption{Corrective effect for $y_i\phi_i x$ under different $a$ and $p$}
\label{ap}
\end{figure}

Further analysis on Step 3 reveals the roles of the parameters $a$ and $p$ in the algorithm. We put emphasis on the weighting coefficient $g(i)^p(2-g(i))$ of $\mathbf{\Phi }$'s $i$th column, which reflects the corrective effect on the previous estimate imposed by the term $y_i\phi_i x$. Fig. \ref{ap} shows how $g(i)^p(2-g(i))$ changes with $y_i\phi_i x$. The higher the amplitude, the greater the corrective effect is imposed by the corresponding $y_i\phi_i x$.

It can be observed from Fig. \ref{ap} that the parameter $p$ selects the region to be corrected. The part of each curve that is on the left half of the figure forces negative $y_i\phi_i x$ above zero and the rest part pushes small positive $y_i\phi_i x$ larger. The parameter $a$ controls the effective corrective range of $y_i\phi_i x$. A small $a$ will result in a large range of $y_i\phi_i x$. The value of $a$ can be optimized according to the value of $p$ and the variance of $y_i\phi_i x+n_i$. For example, although the positive parts of the two curves $a$=2, $p$=2 and $a$=1, $p$=4 are nearly identical, the algorithm with $a$=1, $p$=4 is expected to generate a better performance, as it corrects a larger range of negative $y_i\phi_i x$ than $a$=2, $p$=2.

\section{numerical results}
We carry out several numerical experiments to explore the performance of the SCRs. Here, SCRs are implemented in the following three forms: SCR-1, SCR-2 and SCR-4. These three algorithms, together with BIHT-$l_1$ and BIHT-$l_2$,  are performed and the comparison results are discussed.

Every curve is obtained through averaging the data over 10000 trials. Two kinds of error are calculated to measure the recovery accuracy, the average angular error defined as $\frac{1}{\pi}acos\langle\hat{x},x\rangle$ and the average Hamming error defined as $\frac{1}{M}||sign(\mathbf{\Phi }\hat{x})-y||_0$. Angular error, the main metric,  measures the difference between $\hat{x}$ and $x$, while Hamming error represents the inconsistent part between $sign(\mathbf{\Phi }\hat{x})$ and $y$ and serves as the auxiliary metric.

In all experiments, we set $N=128$ and $K=16$. Measurement matrix $\mathbf{\Phi }$ is assumed to be a Gaussian matrix with independent and identically distributed (i.i.d.) elements. K non-zero elements of signal x are also assumed to be i.i.d. Gaussian. Specifically, we assume $\mathbf{\Phi }\sim N^{M\times N}(0,1)$, $x\sim N^{N\times 1}(0,1)$ and $n\sim N^{M\times 1}(0,\sigma_n^2)$.

\subsection{Accuracy versus Measurement Number}
\begin{figure} [t]
\centering
\includegraphics [width=3.5in]{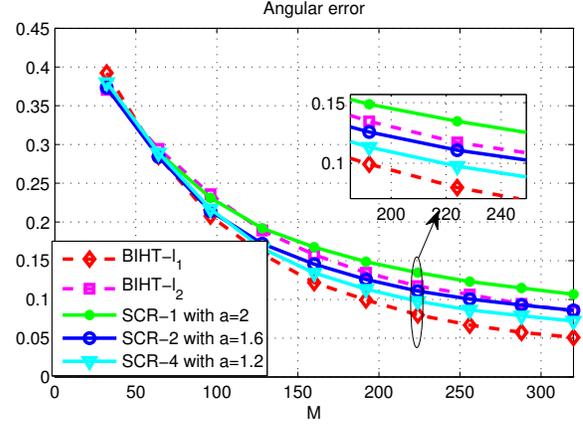}
\caption{Angular error under $\sigma_n^2=0$}
\label{fig4}
\end{figure}
\begin{figure} [t]
\centering
\includegraphics [width=3.5in]{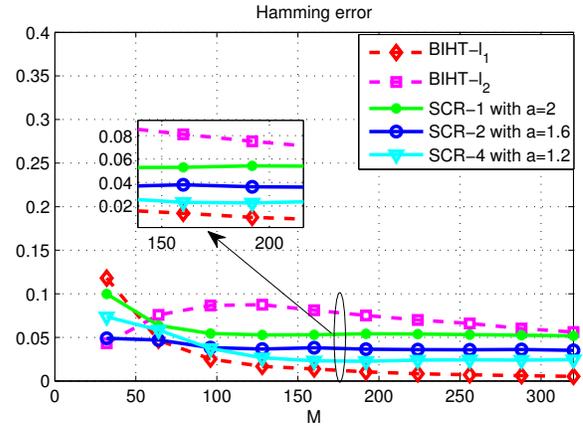}
\caption{Hamming error under $\sigma_n^2=0$}
\label{fig5}
\end{figure}

In the first experiment, we adjust the number of measurements $M$ from $0$ to $320$ and compare the performance of the five algorithms under noise level $\sigma_n^2=0$ and $\sigma_n^2=5$, respectively. The parameter $a$ in each SCR-$p$ algorithm is optimized empirically to achieve the approximately best performance in each scenario.

\begin{figure} [t]
\centering
\includegraphics [width=3.5in]{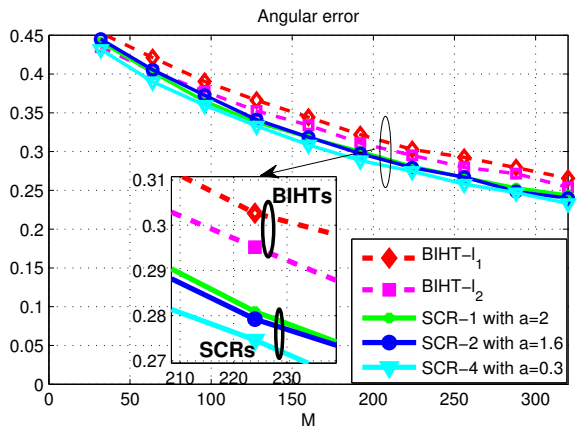}
\caption{Angular error under $\sigma_n^2=5$}
\label{fig6}
\end{figure}

\begin{figure} [t]
\centering
\includegraphics [width=3.5in]{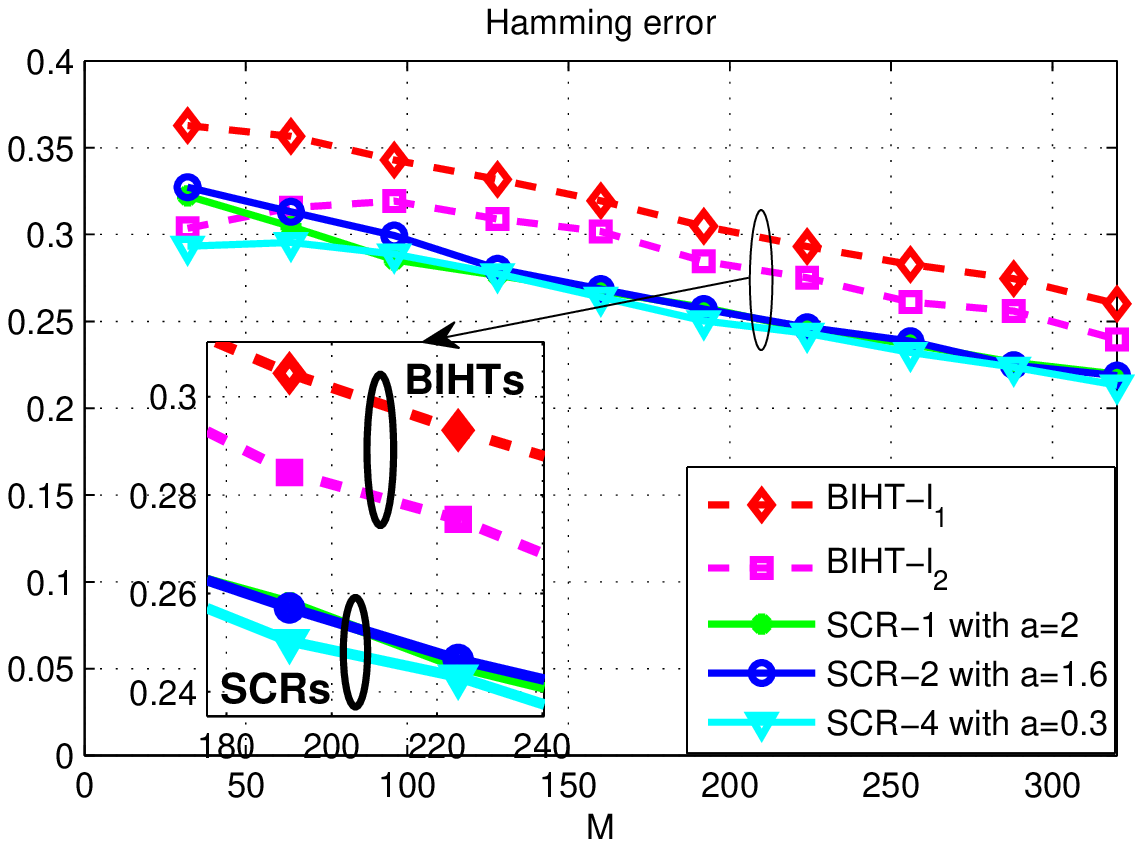}
\caption{Hamming error under $\sigma_n^2=5$}
\label{fig7}
\end{figure}

Fig. \ref{fig4} and Fig. \ref{fig5} present the algorithm comparison under $\sigma_n^2=0$. In noiseless environment, SCR-2 and SCR-4 perform slightly better than BIHT-$l_2$, but all three SCRs are inferior to BIHT-$l_1$. Fortunately, this inferiority may be easily compensated. As is shown in \cite{jacques2011robust}, the estimation accuracy of the BIHTs will continue to improve as long as the number of measurements increases, as the SCRs will do. Unlike the noiseless case, there exists an error floor under noisy environments, no matter how many measurements are used. So the following noisy case bears more importance. As shown in Fig. \ref{fig6} and Fig. \ref{fig7} with noise level $\sigma_n^2=5$, SCR algorithms are superior to BIHT algorithms in both Hamming error and angular error.


\subsection{The Effect of Noise}

\begin{figure} [t]
\centering
\includegraphics [width=3.5in]{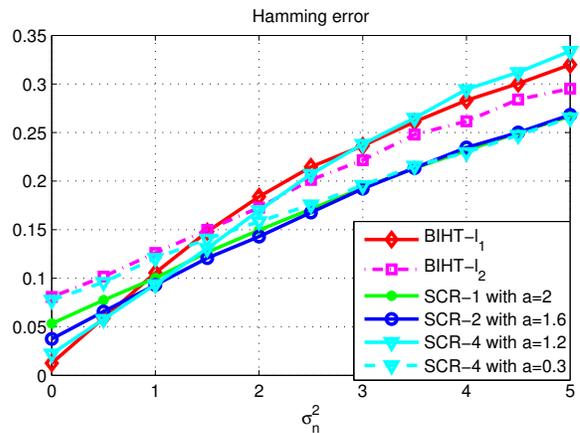}
\caption{Enforcing consistency under noise: Hamming error}
\label{fig9}
\end{figure}
\begin{figure} [t]
\centering
\includegraphics [width=3.5in]{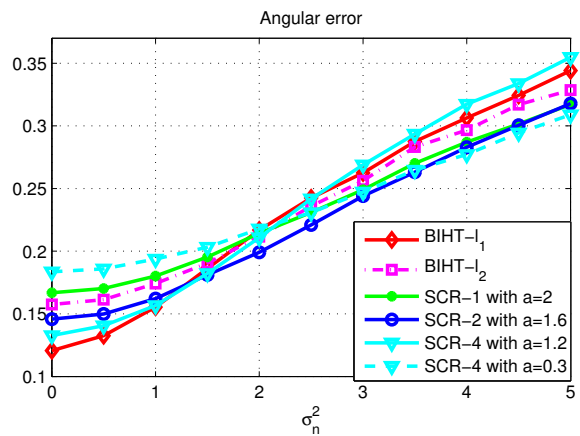}
\caption{Enforcing consistency under noise: angular error}
\label{fig10}
\end{figure}

In this experiment, we fix $M=160$ and adjust the variance of noise in quantized measurements within $\sigma_n^2\in[0,5]$. Fig. \ref{fig9} and Fig. \ref{fig10} show how the performances of SCRs and BIHTs change with the increasing noise level. Their performance  are sorted in high SNR regime (small $\sigma_n^2$) and low SNR regime (large $\sigma_n^2$), as concluded in Table \ref{tabel}. Five algorithms are numbered as: (1)BIHT-$l_1$, (2)BIHT-$l_2$, (3)SCR-1, (4)SCR-2 and (5)SCR-4. The symbol ``$>$" indicates better accuracy performance, i.e., the one on the left is more accurate than the one on the right.

\begin{table}[h]
\caption{Reconstruction Accuracy of Five Algorithms}
\centering
\begin{tabular}{c||c}
\hline
  \hline
Noise level   & Rank according to Hamming error\\
  \hline
high SNR&(1)$>$(5)$>$(4)$>$(3)$>$(2)\\
  \hline
low SNR&(3)$\approx$(4)$\approx$(5)$>$(2)$>$(1)\\
  \hline
  \hline
Noise level&Rank according to Angular error\\
  \hline
high SNR&(1)$>$(5)$>$(4)$>$(2)$>$(3)\\
  \hline
low SNR&(3)$\approx$(4)$\approx$(5)$>$(2)$>$(1)\\
  \hline\hline
\end{tabular}
  \label{tabel}
\end{table}

As expected, the SCR family overwhelmingly surpasses the BIHT family under noisy environments. In addition, each SCR-p exhibits nearly the same performance in low SNR regime. In high SNR regime, BIHT-$l_1$ works best and its performance curve is used as a benchmark. Surprisingly, the performances of the SCRs are no worse than BIHT-$l_2$ and moreover, the larger the order $p$ is used, the closer the performance gap between the SCRs and the benchmark will be, which re-confirms the previous claim that the SCRs are suitable for both high and low SNR regimes.

\subsection{Hamming Error versus Angular Error}

\begin{figure}[t]
\centering
\subfigure[$\sigma_n^2=0$]{\includegraphics[width=1.7in]{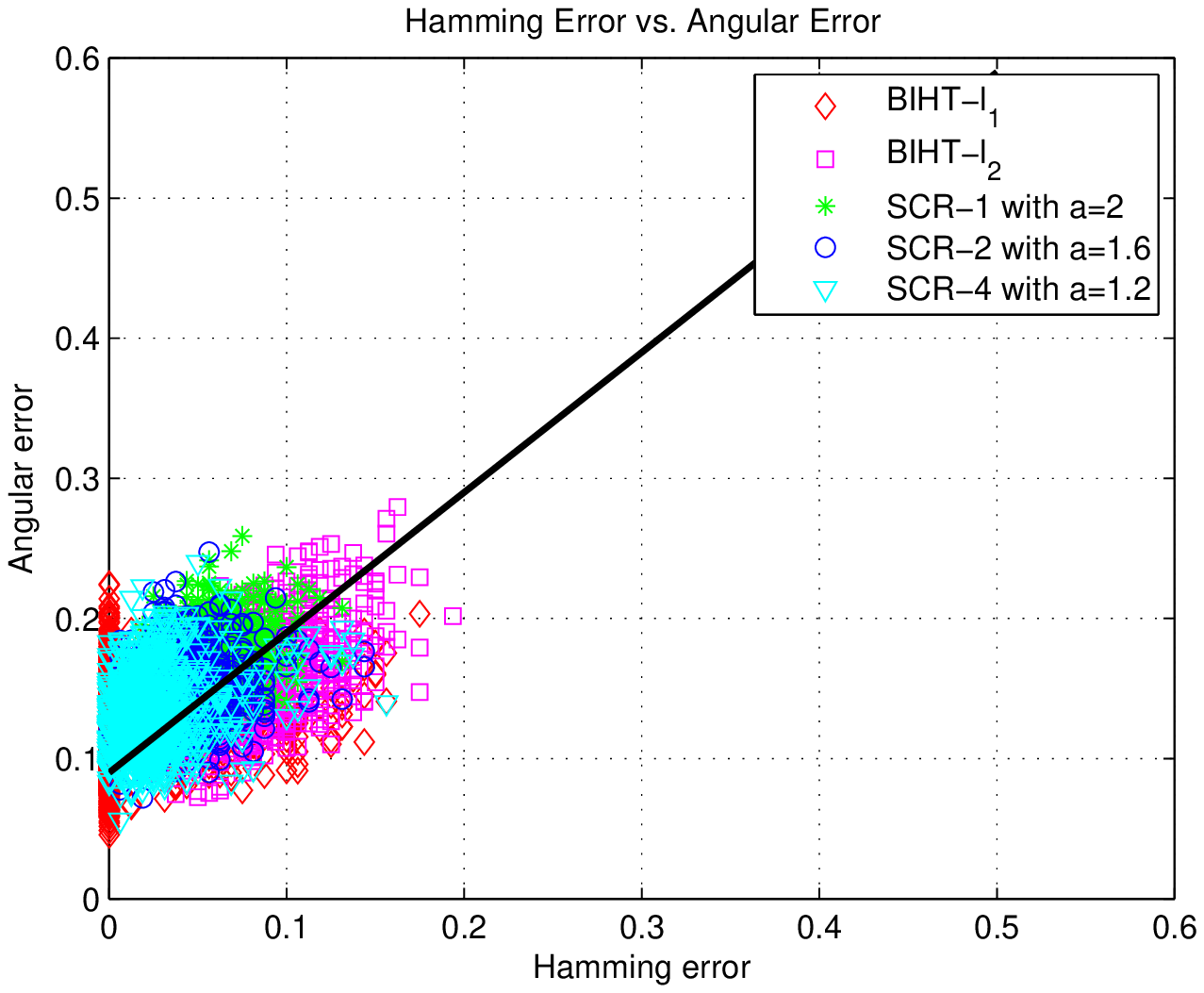}}
\subfigure[$\sigma_n^2=5$]{\includegraphics[width=1.7in]{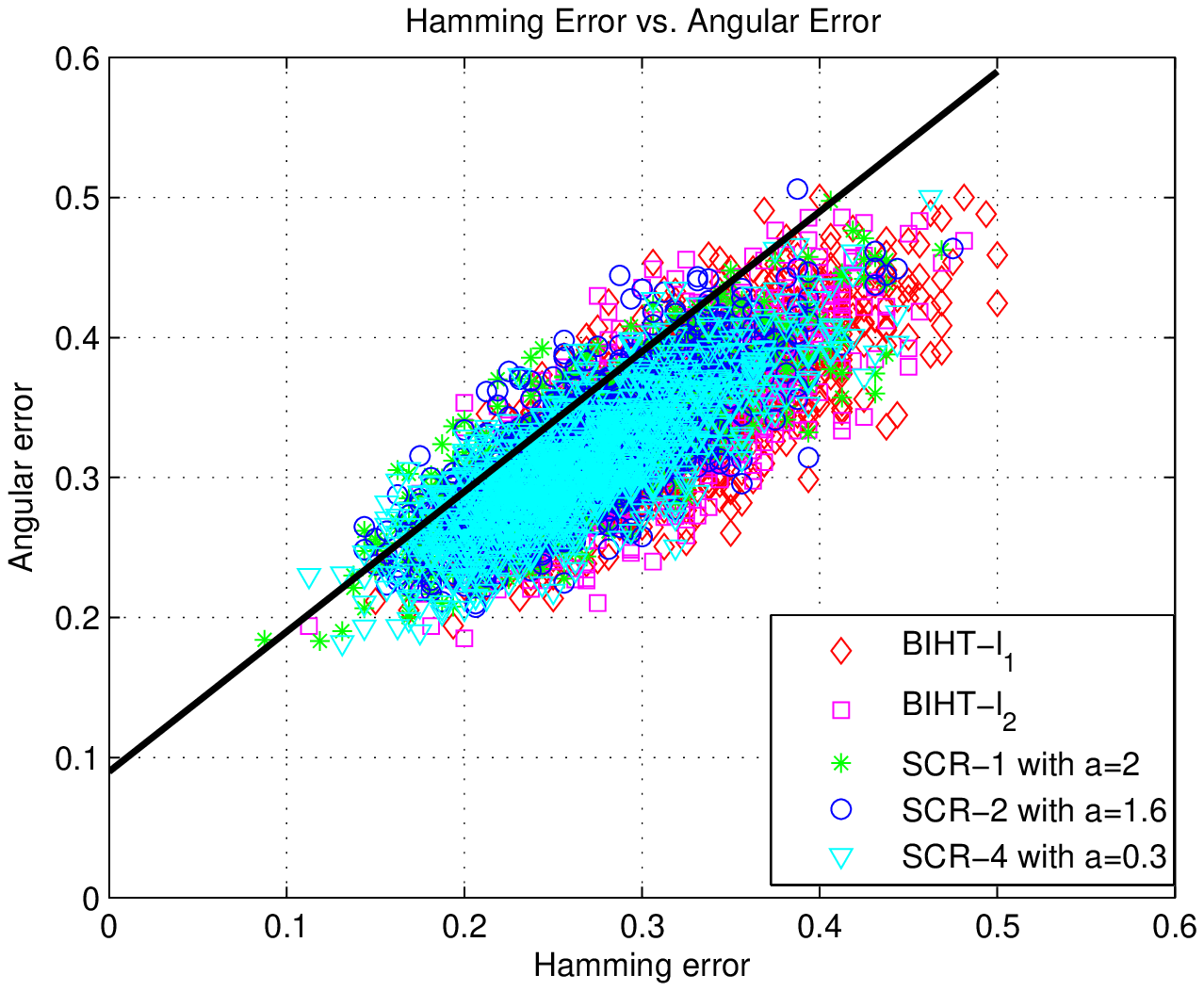}}
\caption{Hamming v.s. angular error}
\label{fig8}
\end{figure}

At last, we explore the relationship between Hamming error and angular error. By fixing $M=160$ and gathering all trial results, we plot Fig. \ref{fig8}(a) for $\sigma_n^2=0$ and Fig. \ref{fig8}(b) for $\sigma_n^2=5$. Fig. \ref{fig8}(a) clearly shows the intermediate performances of  SCRs between BIHT-$l_1$ and BIHT-$l_2$. In Fig. \ref{fig8}(b), the points of SCRs are closer to origin, indicating that SCRs are more robust against noise than BIHTs in both Hamming error and angular error.

\section{Conclusion}
In this paper, we propose a class of recovering algorithms for 1-bit compressive sensing named Soft Consistency Reconstructions (SCRs). Different from the other similar algorithms, the SCR algorithm introduces a new metric to measure consistency between the estimate and binary measurements. The mechanism that makes the SCRs more resistive to noise than the BIHTs is investigated. Experiments verify that the SCRs are superior to the BIHT algorithms in low SNR regime and have comparable performance in high SNR regime.

\section*{Acknowledgment}
This work was supported in part by the National Key Basic Research Program of China (2012CB316104), the National Natural Science Foundation of China(61371094), the National Hi-Tech R\&D Program of China (2012AA121605 and 2014AA01A702), the Zhejiang Provincial Natural Science Foundation of China (LR12F01002, LR12F01001) and the Supporting Program for New Century Excellent Talents in University (NCET-09-0701).

\section*{Appendix}
Step 3 in the SCR algorithm $t=-\frac{1}{2}ap\mathbf{\Phi }^T [y\odot(g)^p\odot(2-g)]$ for each iteration $l$ is a gradient of $\sum_{i=1}^{M}G_a(y_i\phi_i x)^p$.

\begin{proof}
First of all, we have to take the derivative of the scalar function $G_a(t)$
\begin{align}\label{derivationG}
\nonumber
G_a'(t)&=(\frac{2}{e^{at}+1})'\\
\nonumber
&=-\frac{2a\cdot e^{at}}{(e^{at}+1)^2}\\
&=-\frac{a}{2}G_a(t)(2-G_a(t)).
\end{align}

The gradient of $\sum_{i=1}^{M}G_a(y_i\phi_i x)^p$ can then be easily obtained from (\ref{derivationG})
\begin{equation}
\begin{split}
&(\sum_{i=1}^{M}G_a(y_i\phi_i x)^p)'\\
=&\sum_{i=1}^{M}p\cdot G_a(y_i\phi_i x)^{p-1}\cdot(G_a(y_i\phi_i x))'\\
=&\sum_{i=1}^{M}p\cdot G_a(y_i\phi_i x)^{p-1}\cdot(-\frac{a}{2})G_a(y_i\phi_i x)(2-G_a(y_i\phi_i x))\cdot y_i\phi_i^T\\
=&-\frac{1}{2}ap\sum_{i=1}^{M}\phi_i^T\cdot [y_iG_a(y_i\phi_i x)^p(2-G_a(y_i\phi_i x))].
\end{split}
\end{equation}

By making weighted summation of M column vectors $\phi_i^T$s, we conclude that
\begin{equation}
(\sum_{i=1}^{M}G_a(y_i\phi_i x)^p)'=-\frac{1}{2}ap\mathbf{\Phi }^T [y\odot g^p\odot(2-g)]
\end{equation}
where $\odot$ represents element-wise multiplication and g is defined in Step 2 in Algorithm 1.
\end{proof}


\nocite{*}
\bibliographystyle{IEEEtran}
\bibliography{reference}

\end{document}